\documentstyle[12pt]{article}
\def\al{\alpha}

\def\de{\delta}

\def\si{\sigma}

\def\fr#1#2{{{#1} \over {#2}}}

\def\frac#1#2{{\textstyle{{#1}\over {#2}}}}

\def\lsim{\mathrel{\rlap{\lower4pt\hbox{\hskip1pt$\sim$}}
    \raise1pt\hbox{$<$}}}
\def\gsim{\mathrel{\rlap{\lower4pt\hbox{\hskip1pt$\sim$}}
    \raise1pt\hbox{$>$}}}
\def\sqr#1#2{{\vcenter{\vbox{\hrule height.#2pt
         \hbox{\vrule width.#2pt height#1pt \kern#1pt
         \vrule width.#2pt}
         \hrule height.#2pt}}}}

\newcommand{\beq}{\begin{equation}}
\newcommand{\eeq}{\end{equation}}
\newcommand{\bea}{\begin{eqnarray}}
\newcommand{\eea}{\end{eqnarray}}
\newcommand{\rf}[1]{(\ref{#1})}

\renewenvironment{thebibliography}[1]
 { \rm
   \begin{list}{\arabic{enumi}.}
    {\usecounter{enumi} \setlength{\parsep}{0pt}
     \setlength{\itemsep}{3pt} \settowidth{\labelwidth}{#1.}
     \sloppy
    }}{\end{list}}

\begin{document}
\titlepage

\begin{flushright}
{COLBY-94-02\\}
{IUHET 275\\}
\end{flushright}
\vglue 1cm

\begin{center}
{{\bf LONG-TERM EVOLUTION AND REVIVAL STRUCTURE\\
	OF RYDBERG WAVE PACKETS
\\}
\vglue 1.0cm
{Robert Bluhm$^a$ and V. Alan Kosteleck\'y$^b$\\}
\bigskip
{\it $^a$Physics Department\\}
\medskip
{\it Colby College\\}
\medskip
{\it Waterville, ME 04901, U.S.A\\}
\bigskip
{\it $^b$Physics Department\\}
\medskip
{\it Indiana University\\}
\medskip
{\it Bloomington, IN 47405, U.S.A.\\}

}
\vglue 0.8cm

\end{center}

{\rightskip=3pc\leftskip=3pc\noindent
It is known that, after formation,
a Rydberg wave packet undergoes a series of collapses
and revivals within a time period called the revival time,
$t_{\rm rev}$,
at the end of which it is close to its original shape.
We study the behavior of Rydberg wave packets
on time scales much greater than
$t_{\rm rev}$.
We show that after a few revival cycles the wave packet
ceases to reform at multiples of the revival time.
Instead,
a new series of collapses and revivals commences,
culminating after a time period $t_{\rm sr} \gg t_{\rm rev}$
with the formation of a wave packet that more closely
resembles the initial packet than does the full
revival at time $t_{\rm rev}$.
Furthermore,
at times that are rational fractions of $t_{\rm sr}$,
the square of the autocorrelation function exhibits
large peaks with periodicities that can be expressed
as fractions of the revival time $t_{\rm rev}$.
These periodicities indicate a new type of fractional
revival occurring for times much greater than $t_{\rm rev}$.
A theoretical explanation of these effects is outlined.

}

\vskip 1truein
\centerline{\it Published in Physics Letters A
{\bf 200}, 308 (1995)}

\vfill
\newpage

\baselineskip=20pt

An electron wave packet in a Rydberg atom
provides a physical system in which to explore the
interface between classical and quantum mechanics.
When a Rydberg atom is excited by a short laser pulse,
a state is created that has classical behavior for a limited time
\cite{ps,az}.
Depending on the excitation scheme,
either a radially localized wave packet
in an eigenstate of angular momentum
\cite{tenWolde,yeazell,meacher}
or a packet localized in the angular coordinates
\cite{angular,gaeta}
is produced.
Either a radial or a circular wave packet
initially oscillates with the classical keplerian period
$T_{\rm cl} = 2 \pi {\bar n}^3$,
where $\bar n$ is the central value of the principal
quantum numbers excited in the packet.
However,
after a finite number of classical orbits
quantum-interference phenomena appear
and the wave packet collapses.

On a time scale large compared to the classical period $T_{\rm cl}$,
a Rydberg wave packet passes through a sequence
of fractional and full revivals
\cite{ps,az,ap,nau1}.
A full revival occurs at a time\footnote{
Note that $t_{\rm rev}$ is often defined as half the
full revival time,
since this is the time at which a full wave packet
reforms for the first time,
though with a phase difference relative to the
classical motion.
We adopt the notation of
Ref.\ \cite{ap}
for labeling the fractional revivals and use
their definition of $t_{\rm rev}$.}
$t_{\rm rev} = \frac {2 {\bar n}} 3 T_{\rm cl}$,
when a packet close to the original shape reappears.
The fractional revivals occur earlier than this,
at times
that are rational fractions of the revival time $t_{\rm rev}$.
They correspond to the formation of macroscopically
distinct subsidiary wave packets that oscillate with periodicity that
is a fraction of the classical orbital period $T_{\rm cl}$
\cite{ap}.
Recent experiments have detected fractional revivals with
periodicities as small as $T \approx \fr 1 7 \, T_{\rm cl}$
\cite{wals}.

In this letter,
we examine the time evolution and revival structure
of both radial and circular Rydberg wave packets
on time scales much greater than the revival time $t_{\rm rev}$.
A new system of full and fractional revivals is uncovered,
with structure different from that of the usual fractional revivals.

In Ref.\ \cite{ps},
long-term revivals at $t= {\bar n}^2 T_{\rm cl}$ and
$t= {\bar n}^3 T_{\rm cl}$ were found.
They were obtained by expanding the energy in a Taylor series
in $n$ through finite order
and determining the times at which the time-dependent
phase in the wave function is
an integer multiple of $2 \pi$.
Ref.\ \cite{peres}
generalized this approach and found a hierarchy
of recurrence times for full revivals of the
wave packet for $t \gg t_{\rm rev}$.

Our approach here is different.
We do not look for commensurability of the terms in
the phase with integer multiples of $2 \pi$.
Rather,
we show that at certain times greater than $t_{\rm rev}$
the wave function can be expanded in a
set of subsidiary waves.
This method is suitable for studying
fractional revivals
\cite{ap}
as well as full revivals.
It also reveals the presence of a long-term
full revival occurring earlier than those found previously
and not among the hierarchy of levels noted in
Ref.\ \cite{peres}.

A simple way of demonstrating
the formation of the new fractional and full revivals
is to construct the
autocorrelation function of the
time-evolved wave function with the initial packet.
At the various fractional-revival times,
the absolute square of the autocorrelation function
exhibits periodicities reflecting
the presence of underlying subsidiary wave packets.
Furthermore,
the ionization signal in a pump-probe experiment
should display the same periodicities as the
autocorrelation function.
Thus,
the primary features of our results
relevant for experiments on Rydberg wave packets can be
exhibited directly in a graph of the autocorrelation
function as a function of time.
For this reason,
we first present plots
of the autocorrelation function
for times greater than $t_{\rm rev}$.
We use these to indicate the structure of the new revivals.
We then outline the theoretical explanation of the effects
and suggest experiments to detect them.

Provided the eigenenergies are independent of the
angular-momentum quantum numbers,
it can be shown that
the autocorrelation function for a given radial
packet is the same as that for
a related circular packet.
For example,
in the case of hydrogen the expression
for the modulus squared of the autocorrelation function
is
\beq
| A(t) |^2 = {\Bigm| \sum_n \bigm| c_n
\bigm|^2 e^{-i E_{n} t} \Bigm|}^2
\quad
\label{auto}
\eeq
for both types of wave packet.
Here, $c_n = \left< \Psi (0) \vert \Psi_n \right>$,
and $\Psi_n$ is an energy eigenstate with principal
quantum number $n$.

We consider Rydberg wave packets of hydrogen for the case where the
excitation spectrum is strongly peaked around a central
value $\bar n$.
For simplicity in the present work,
we take a normalized gaussian
distribution of width $\si$ for the modulus squared
of the coefficients,
$| c_n |^2$.
However,
the results we obtain hold equally well for
asymmetric excitation distributions,
such as those occurring in a description in
terms of radial squeezed states
\cite{rss}.
Our results also can be generalized to the case of noninteger
values of $\bar n$,
including thereby the effects of quantum defects.
This means, for example,
that our results apply equally well to alkali-metal atoms.
Experimentally,
the value of $\si$ is set by the exciting laser pulse.
In what follows,
the values chosen for $\si$ correspond to a laser-pulse duration
that is effective in displaying the new revival structure.

The time-dependent wave function for the Rydberg wave packet
may be written as an expansion in terms of eigenstates
of hydrogen
\beq
\Psi ({\vec r},t) = \sum_{k=-\infty}^{\infty} c_k
\varphi_k({\vec r}) \exp \left( -i E_n t \right)
\quad ,
\label{psi}
\eeq
where $k = n - {\bar n}$, $\bar n$ is assumed large,
and $\varphi_k ({\vec r})$ is a hydrogenic wave function.
Since we are taking a distribution strongly
centered around $\bar n$,
we may expand $E_n$ around $\bar n$:
\beq
\Psi ({\vec r},t) = \sum_{k=-\infty}^{\infty} c_k
\varphi_k({\vec r}) \exp \left[ -2 \pi i
\left( \frac {kt} {T_{\rm cl}} -  \frac {k^2 t} {t_{\rm rev}}
+ \frac {k^3 t} {t_{\rm sr}}  -\ldots +\ldots\right) \right]
\quad .
\label{psi3rd}
\eeq
In this equation,
we have introduced three distinct time scales
$T_{\rm cl}$, $t_{\rm rev}$, and $t_{\rm sr}$
that naturally appear in the first three orders in
the expansion.
The first two are the usual time scales
relevant in the description of the conventional revival structure.
The smallest time scale, $T_{\rm cl}$, is the classical period
for an electron in a keplerian orbit.
If this term alone appeared in the expansion,
the energies would be equally spaced
and the wave packet would undergo simple harmonic motion
without changing shape.
The second term defines the revival time scale,
$t_{\rm rev}$,
and is responsible for the collapse
and fractional/full revivals of the wave packet
\cite{ap}.

The third-order term is included
because in this paper we are interested in times much greater
than the revival time.
This term defines a new time scale,
which we refer to as the superrevival time
$t_{\rm sr}$.
In terms of the revival time,
it is given by
$t_{\rm sr} = \frac {3 {\bar n}} 4 t_{\rm rev}$.
For typical values of $\bar n$,
$t_{\rm sr}$ is between one and two orders of magnitude
greater than the revival time $t_{\rm rev}$.
However,
this is still several orders of magnitude
less than the lifetime of the excited Rydberg atom.
It is therefore reasonable
to examine the behavior of the wave packet and determine the
effects of the third-order term on
the revival structure up to times of order $t_{\rm sr}$.

With the gaussian distribution for $|c_n|^2$ discussed above,
we can evaluate the absolute square of the autocorrelation
function $| A(t) |^2$ directly from
Eq.\ \rf{auto}
and plot the result.
We begin with an example involving a circular wave packet
that serves to illustrate some
key features of the new revival structure.
Subsequently,
we examine a radial wave packet with a lower value of $\bar n$
of a type that could be experimentally created and observed.

Consider the circular wave packet treated in
Ref.\ \cite{gaeta},
with ${\bar n} = 320$,
$l = m = n-1$,
and $\si = 2.5$,
which displays a clear sequence
of conventional fractional/full revivals
on the revival timescale $t_{\rm rev}\simeq 1.06$ $\mu$sec.
Figure 1 shows the absolute square of the autocorrelation function
for a Rydberg wave packet with ${\bar n} = 320$
for a total time period large compared to the revival time.
In this case,
$t_{\rm sr}\simeq 255$ $\mu$sec,
and we consider times up to just beyond $\frac 1 6 t_{\rm sr}$.
Figure 1a shows the behavior of the wave packet for
approximately the first 9 $\mu$sec.
The full revival appears at $t_{\rm rev} \simeq 1.06$ $\mu$sec,
and fractional revivals are evident at fractions of this value.
The largest revival appears at $t = \frac 1 2 t_{\rm rev}$,
where the wave function has reformed into a single wave packet,
but at a time that is out of phase with a classically
propagated distribution
\cite{gaeta}.

As can be seen from the figure,
there are four or five full revival cycles,
with peaks gradually decreasing in size and losing
the canonical periodicity.
However, after about 7 $\mu$sec
a new structure with a noncanonical periodicity
is visible in the autocorrelation function.
Similar structures reappear later,
at times near 14, 21, 28, and 42.5 $\mu$sec
in Figures 1b, 1c, 1d, and 1e,
respectively.
The amplitudes and periodicities
of these structures change with time.
At $t \simeq 42.5$ $\mu$sec,
a series of peaks appears with maximum amplitude
greater than that of the full revival at $t = t_{\rm rev}$.

The features of this illustrative example
can be predicted theoretically.
We have proved that
at times $t \approx \frac 1 q t_{\rm sr}$,
where $q$ must be an integer multiple of 3,
the wave packet can be written as a sum of
macroscopically distinct wave packets.
The proofs are rather technical and are presented elsewhere
\cite{sr2}.
Explicitly,
we find that near times $t \approx \frac 1 q t_{\rm sr}$
we may write
\beq
\Psi ({\vec r},t) = \sum_{s=0}^{l-1} b_s
\psi_{\rm cl} ({\vec r},t + \frac {s \al} l T_{\rm cl} )
\quad ,
\label{expans}
\eeq
where the macroscopically distinct wave packets
are defined by
\beq
\psi_{\rm cl} ({\vec r},t) =
\sum_{k=-\infty}^{\infty} c_k
\varphi_k({\vec r})
\exp \left[ -2 \pi i
\left( \frac {kt} {T_{\rm cl}} \right) \right]
\quad
\label{psicl}
\eeq
and where their weights $b_s$ in the sum \rf{expans} are
given by
\beq
b_s = \fr 1 l
\sum_{k^\prime =0}^{l-1}
\exp \left[ 2 \pi i
\left(
\fr {\al s} l k^\prime + \fr {3\bar n} {4q} k^{\prime 2}
- \fr 1 q k^{\prime 3}
\right) \right]
\quad .
\label{bs}
\eeq
The quantities
$\al$ and $l$ are integer constants that depend on $q$
and $\bar n$:
\beq
\al = \fr{ 2 \bar n}N~~,~~~~
l =\cases{q&if~~$q/9 \ne 0~~ ({\rm mod}~1)~~$,\cr
        q/3&if~~$q/9 = 0~~ ({\rm mod}~1)~~$,\cr }
\quad
\label{ints}
\eeq
where $N$ is an integer consisting of the product
of all factors of $2\bar n$ that are also factors of $l$.

We have shown that the properties of the $b_s$ coefficients
are such that the autocorrelation function is periodic,
with a period\footnote{
The actual periodicity is
$T = \frac 3 q t_{\rm rev} + \frac u v T_{\rm cl}$,
where $u$ and $v$ are integers that
depend on $\bar n$ and $q$.
Since $T_{\rm cl} \ll t_{\rm rev}$,
we write $T \approx \frac 3 q t_{\rm rev}$.}
$T \approx \frac 3 q t_{\rm rev}$.
These periodicities are on a much greater time scale than the
usual fractional revivals,
which have periodicities that are fractions of $T_{\rm cl}$.
Moreover,
for special values of $q$
the number of nonzero $b_s$ coefficients is small and
one coefficient dominates.
This means that the wave function
is mostly concentrated in one packet,
which explains the peaks in the autocorrelation function.
In special cases,
only one nonzero $b_s$ value exists.
This first happens
when $t \approx \frac 1 6 t_{\rm sr}$,
so that $q=6$.
It corresponds to the formation of a single packet,
i.e., a full superrevival,
with periodicity $T \approx \frac 1 2 t_{\rm rev}$
in the autocorrelation function.
For the fractional times $t \approx \frac 1 q t_{\rm sr}$
with $q>6$,
we find that more than one $b_s$ coefficient is nonzero,
corresponding to the formation of distinct subsidiary
wave packets at times $t \gg t_{\rm rev}$.
We refer to these as fractional superrevivals.

Since the $b_s$ coefficients in
\rf{expans}
have moduli that are generally not equal,
this indicates that the fractional superrevivals
consist of an unequally weighted superposition of
subsidiary wave packets,
which is a feature that is not true for the usual
fractional revivals.
Moreover,
after the formation of a fractional superrevival
consisting of several subsidiary wave packets,
it often happens that the subsidiary packets quickly
evolve into a configuration where one of them is much
greater than the others.
The dominant wave packet in this case can again resemble
the initial wave packet more closely than the full revival
does at time $t_{\rm rev}$.
In this way,
large peaks can form in the autocorrelation function
near the fractional superrevival times.

We demonstrate these results
for the illustrative example with ${\bar n} = 320$.
In this case,
the theory predicts
the first full superrevival appearing at
$t \simeq 42.5$ $\mu$sec,
with an autocorrelation-function periodicity
$T \simeq 530$ nsec.
This agrees with the structure seen in Figure 1e.
For $q = 36$, $18$, $12$, and $9$,
fractional superrevivals are predicted to appear,
since for these values of $q$ more than
one of the $b_s$ coefficients is nonzero.
The corresponding
times at which structures should be seen
are $t \simeq 7.08$, $14.2$, $21.2$, and
$28.3$ $\mu$sec,
and they should have autocorrelation-function
periodicities $T \approx \frac 1 {12} t_{\rm rev}$,
$\frac 1 {6} t_{\rm rev}$,
$\frac 1 {4} t_{\rm rev}$,
and $\frac 1 {3} t_{\rm rev}$,
respectively.
These predictions agree with
the structures seen in Figs. 1a -- d.

To obtain further insight into the formation of a superrevival,
we can examine the cross-sectional view
of a circular wave packet at various times.
Figure 2 shows a cross-sectional slice of the circular wave
packet with ${\bar n} = 320$.
We view the wave packets in the plane of the orbit and
plot a circular slice $\Psi (\phi)$ at a fixed radius
given by the expectation value $\left< r \right> =
\frac 1 2 {\bar n} (2{\bar n} + 1)$.
Figure 2a shows the initial wave packet at time $t=0$.
Figure 2b shows the wave packet
at the canonical revival time $t \approx
t_{\rm rev}$ when the autocorrelation function is a local maximum.
As can be observed,
the wave packet has indeed reformed
and resembles the original wave,
but it is asymmetric
and several smaller subsidiary wave packets also appear.
Figure 2c shows the full superrevival
at $t \approx \frac 1 6 t_{\rm sr}$,
at which point only one $b_s$ coefficient in
Eq.\ \rf{expans} is nonzero.
In comparing Figures 2b and 2c,
the superrevival wave packet is seen to resemble
more closely the initial wave packet of Figure 2a
than does the canonical revival packet.

We chose ${\bar n} = 320$ as an illustrative
example both for purposes of comparison with
previous results for $t \le t_{\rm rev}$
and because
the full and fractional superrevivals are more evident
for larger $\bar n$.
However,
similar effects can be observed for Rydberg wave packets
with relatively small values of $\bar n$.

Figure 3 shows the square of the autocorrelation
function for hydrogen with ${\bar n} = 48$ and $\si = 1.5$.
As remarked above,
the autocorrelation function for hydrogen is the same
for both radial and circular wave packets.
Thus,
Figure 3 represents the autocorrelation function
of a radial Rydberg wave packet that has been excited by
a short laser pulse centered on the value ${\bar n} = 48$.
In this case,
the full revival is at $t \approx t_{\rm rev} \simeq 0.538$ nsec.
For times greater than this,
one can observe a fractional superrevival at $t \approx \frac 1 {12}
t_{\rm sr} \simeq 1.61$ nsec
with autocorrelation periodicity $T \approx \frac 1 4 t_{\rm rev}$
and a full superrevival at
$t \approx \frac 1 6 t_{\rm sr} \simeq 3.23$ nsec
with autocorrelation periodicity $T \approx \frac 1 2 t_{\rm rev}$.
The size of the peak in the autocorrelation function shows
that the superrevival resembles the initial wave packet
more closely than does
the revival wave packet at $t \approx t_{\rm rev}$.

We have proved that similar results hold when
quantum defects are present in the energies.
Therefore,
we expect superrevivals to occur
for wave packets in alkali-metal atoms as well as in hydrogen
\cite{bkrc}.
In alkali-metal atoms,
the associated characteristic times depend on ${\bar n}^\ast
= {\bar n} - \de (l)$ instead of $\bar n$,
where $\de (l)$ is the quantum defect.

Figure 3 indicates it is
likely an experiment can be performed to detect
the full and fractional superrevivals discussed in this paper.
One possibility is
to use the pump-probe time-delayed photoionization
method of detection
for radial Rydberg wave packets excited in
alkali-metal atoms with
${\bar n} \approx 45$ -- $50$.
Single-photon absorption would produce a packet
with localization purely in the radial coordinate
and with the angular structure of a p state.
The procedure is experimentally feasible,
provided a delay line of 3 -- 4 nsec
is installed in the apparatus.
In such an experiment,
the fractional superrevival at $\fr 1 {12} t_{\rm sr}$,
with periodicity $T \approx \fr 1 4 t_{\rm rev}$,
could be observed with a delay line of approximately
1.5 nsec.
For even smaller values of $\bar n$,
the required delay times can be reduced below 1 nsec.
With ${\bar n} \simeq 36$,
for example,
the full/fractional superrevivals could be detected
with delay lines used currently in experiments.

\vglue 0.3cm

We enjoyed conversations with Charlie Conover
and Duncan Tate.
R.B. thanks Colby College for a Science Division Grant
and the Theory Group at Indiana University for its
hospitality.

\vfill\eject

\baselineskip=16pt

\begin{description}

\item[{\rm Fig.\ 1:}]
The absolute square of the autocorrelation function for a
Rydberg wave packet with ${\bar n} = 320$ and $\si = 2.5$
is plotted as a function of time in microseconds.
The value $\si=2.5$ for a wave packet with $\bar n= 320$
corresponds to excitation with a 160 picosecond laser pulse.
(a) $0\le t\le 9$ $\mu$sec,
(b) $9\le t\le 18$ $\mu$sec,
(c) $18\le t\le 27$ $\mu$sec,
(d) $27\le t\le 36$ $\mu$sec,
(e) $36\le t\le 45$ $\mu$sec.

\item[{\rm Fig.\ 2:}]
Unnormalized circular wave packets with ${\bar n} = 320$
and $\si = 2.5$.
Cross-sectional slices of the wave packet in the plane
of the orbit and for $r = \left< r \right> =
\frac 1 2 {\bar n} (2{\bar n} + 1)$ are plotted as
a function of the azimuthal angle $\phi$ in radians.
(a) $t=0$,
(b) $t \approx t_{\rm rev}$,
(c) $t \approx \frac 1 6 t_{\rm sr}$.

\item[{\rm Fig.\ 3:}]
The absolute square of the autocorrelation function for a
Rydberg wave packet with ${\bar n} = 48$ and $\si = 1.5$
is plotted as a function of time in nanoseconds.
The value $\si=1.5$ for a wave packet with $\bar n= 48$
corresponds to excitation with a 900 femtosecond laser pulse.

\end{description}

\vfill
\eject

\end{document}